\documentclass[twocolumn,preprintnumbers,superscriptaddress,amsmath,amssymb]{revtex4}

\usepackage{graphicx}
\usepackage{dcolumn}
\usepackage{bm}
\usepackage{color}

\begin{document}
\begin{center}
Published in Eur. Phys. J. D (2010), DOI:~10.1140/epjd/e2010-00004-1 
\end{center}

\title{Spectroscopy of electron-induced fluorescence in organic liquid
scintillators}

\author{T. Marrod\'an Undagoitia} 
\altaffiliation{Corresponding author. Fax: +41 44 63 55704,
Telephone Number: +41 44 63 56686.} 
\email{marrodan@physik.uzh.ch}
\affiliation
{Physik-Department, Technische Universit\"at M\"unchen,\\
James-Franck-Str., 85748 Garching, Germany}
\affiliation
{Physik-Institut, Universit\"at Z\"urich,\\
Winterthurerstr. 190, 8050 Z\"urich, Switzerland}
\author{F. von Feilitzsch}
\author{L. Oberauer}
\author{W. Potzel} 
\author{A. Ulrich}
\author{J. Winter} 
\author{M. Wurm} 
\affiliation
{Physik-Department, Technische Universit\"at M\"unchen,\\
James-Franck-Str., 85748 Garching, Germany}

\begin{abstract}
Emission spectra of several organic liquid-scintillator mixtures which
are relevant for the proposed LENA detector have been measured by
exciting the medium with electrons of $\sim 10$\,keV. The results are
compared with spectra resulting from ultraviolet light
excitation. Good agreement between spectra measured by both methods has
been found.
\end{abstract}

\maketitle

\begin{center}
\end{center}

\section{Introduction}

A new experimental method has been used to measure emission 
spectra of organic liquid scintillators. Commonly, such
measurements are performed by exciting the medium with ultraviolet
(UV) light. However, in an actual particle-astrophysics experiment
charged particles excite the detection medium by depositing their kinetic
energy. In this work, two methods have been used for the excitation of
the scintillator: an UV-light emitter (a Deuterium lamp) and an
electron beam of about 10\,keV energy. The experimental setups for
both methods are described and the resulting spectra are compared.
Electron-beam induced fluorescence has previously been used to study
nitrogen fluorescence in the atmosphere\,\cite{Ave:2007xh}. Such
processes are important for experiments which study cosmic-ray induced
extended air showers via their fluorescence-light emission.
For an optimal choice of a liquid-scintillator detection medium for a
particle-physics detector, several optical parameters have to be taken
into account. One of them is the scintillation emission-spectrum as it
has an impact on the light propagation through the liquid. Absorption
and scattering processes are wavelength dependent. In organic liquids
absorption and scattering lengths increase with
wavelength\,\cite{Alimonti:2000wj}. Additionally, the photo-cathode of
the light sensors in the experiment have a wavelength-dependent
sensitivity. For these reasons, it is of interest to measure the
emission spectra of several scintillator candidates. 

The spectroscopic measurements described in the present paper have
been performed within feasibility studies for the proposed LENA
({\bfseries L}ow {\bfseries E}nergy {\bfseries N}eutrino {\bfseries
A}stronomy)
detector\,\cite{MarrodanUndagoitia:2006re}\cite{Undagoitia:2005uu}\cite{PhD_TMarrodan}.  LENA is a large-volume (50\,kt) observatory based on the 
liquid-scintillator
technology. It aims at the investigation of a variety of topics in
astrophysics, geophysics and particle physics. The current detector
design foresees a cylindrical shape of 100\,m height and 30\,m
diameter placed in vertical position. Light produced in the central
axis of the detector has to reach the photo-sensors which are placed
at the walls of the detector. Therefore, an attenuation length of at
least 10\,m is required\,\cite{Undagoitia:2005uu}. For this detector, 
the scintillation emission spectrum of the medium plays a major role 
as high transparency is necessary.

\section{Scintillation processes}

An organic liquid scintillator usually consists of a solvent medium
and one or more wavelength-shifters. The kinetic energy of charged
particles crossing the medium is deposited mainly in the solvent
molecules and excites the electrons (in particular the
$\pi$-electrons) of benzene-ring structures of the solvent. The
$\pi$-electrons are delocalized and only weakly bonded. The
excitation energy is rapidly transferred to the solute, usually by
dipole-dipole interaction\,\cite{Birk}\cite{CBuck_PhD}. The efficiency
of this transfer depends on the overlap of the emission spectrum of
the solvent and the absorption spectrum of the solute. The
de-excitation of the $\pi$-electron in the solute molecule leads to
the emission of observable light originating from the transition
between the first excited spin-singlet state and one of the
vibrational levels of the ground state. A typical emission spectrum of
such materials is a band of $50 - 100$\,nm width resulting from
overlapping peaks. The relative intensities of the transitions to the
different vibrational levels are given by the Franck-Condon
factors\,\cite{Franck1926}\cite{Condon1926}. Depending on the
interaction of the solvent with the solute, the spectra might be broadened
and/or slightly shifted in wavelength.

\section{The spectrometer}
\label{Spectrometer}
The spectrometer used is of the type Ocean Optics HR2000\-CG-UV-NIR
which is a medium-resolution miniature device\,\cite{OceanOp_Web}. It
covers a wavelength range from 200 to 1100\,nm with a measured
resolution of 0.53\,nm (full width at half maximum). The light is
introduced into the spectrometer via a quartz light-guide (600\,$\mu$m
in diameter). The module utilizes a 5\,$\mu$m entrance slit.  Inside
the spectrometer, the light is dispersed by a blaze grating which has
a groove-spacing density of 300\,lines/mm. The light is finally
recorded by a Sony ILX511 linear CCD (charge-coupled device)
array. Higher diffraction orders of the grating are blocked by edge
filters. For the readout of the detector, the CCD data is extracted by
a USB cable which connects the spectrometer to a computer. The output
data can be analyzed offline with a custom-made program, e.g. with
ROOT\,\cite{ROOT}.

\begin{figure}[h]
  \begin{center}
   \includegraphics[width=0.495\textwidth]{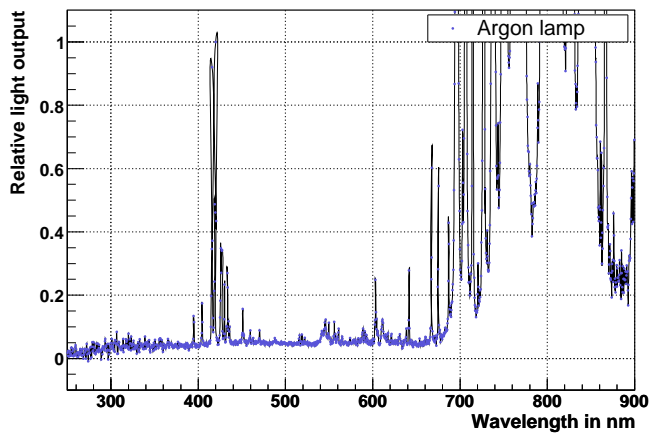}
   \caption[SpectrometerCorrCurv]{Measured spectrum of the argon
   lamp.\label{ArgonSpectrum}}
  \end{center}	
\end{figure}

The spectrometer has to be calibrated in energy and the spectra have
to be corrected for the intensity response at different
wavelengths. For the calibration of the wavelength, a low-pressure
discharge argon-lamp was used. Figure\,\ref{ArgonSpectrum} shows the
measured spectrum.
The lines correspond to transitions of excited states of Ar atoms. The
positions of distinct lines in the spectrum are tabulated. Several
lines between $\sim350$ and 650\,nm were used. The calibration showed
a shift of the argon spectrum by $1.22\pm0.04$\,nm to higher values of
the wavelength. This shift has been corrected for in the data
analysis.

\begin{figure}[h]
  \begin{center}
   \includegraphics[width=0.495\textwidth]{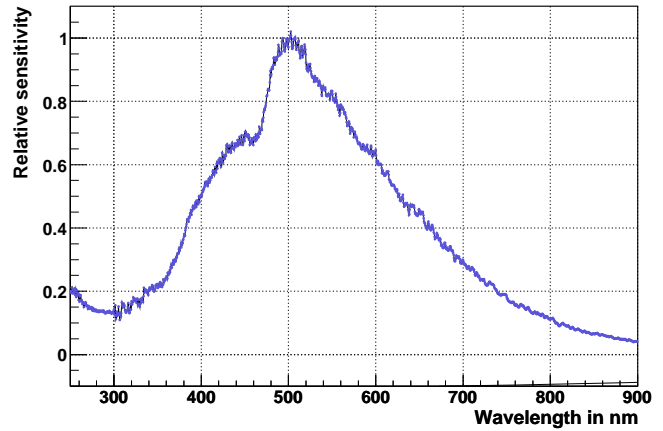}
   \caption[SpectrometerCorrCurv]{Measured response of the
   spectrometer at different wavelenghts. The detector is most
   sensitive at 500\,nm, while e.g. at 300\,nm the efficiency drops to
   $\sim$15\% of the maximum.\label{SpectrometerCorrCurv}}
  \end{center}	
\end{figure}

The wavelength dependence of the sensitivity was measured using a
calibrated halogen lamp (LOT-Oriel\,\cite{LOT}). The lamp has a known
spectral emission when it is powered with 6.60\,A (16.1\,V) and the
spectrum is measured at 70\,cm distance. The spectrometer was placed
at this distance and the measured spectrum was compared to the one
given by the manufacturer of the halogen lamp. In
figure\,\ref{SpectrometerCorrCurv}, 
the relative spectral response of the spectrometer is shown.

\section{Experimental setups}
\label{setups}

\begin{figure}[h]
  \begin{center}
   \includegraphics[width=0.46\textwidth]{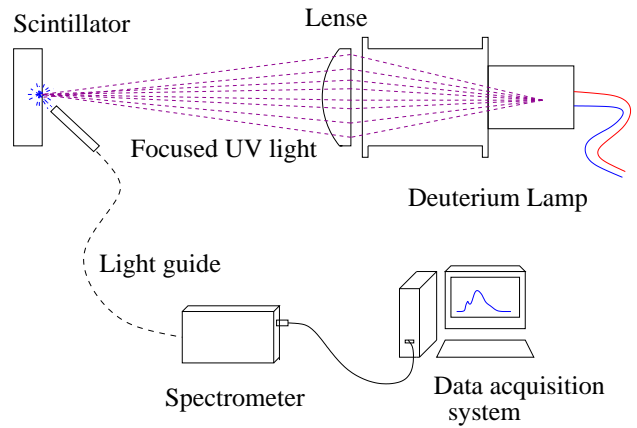}
   \caption[Picture of the setup]{Setup of the spectral measurement of
   liquid scintillators when exciting the sample with UV-light
   of a deuterium lamp.\label{SetUp_UV_Photo}}
  \end{center}	
\end{figure}

Spectra were recorded by exciting the scintillator with an
UV-light source, a deuterium lamp. The lamp is a low-pressure
gas-discharge light source.  The optical output-window of the lamp is
made of magnesium fluoride in order to prevent UV-light
absorption. The light emitted by the lamp was collimated
and focused onto the scintillator
sample by a quartz lens. Figure\,\ref{SetUp_UV_Photo} shows a schematic 
drawing of the setup.
The container for the scintillator sample was made out of black PTFE
(Polytetrafluoroethylene) material which is known not to react with
liquid scintillators. The black color was chosen to suppress light
reflexions. The container is a cylinder with an inner diameter of
2.5\,cm and 1.0\,cm length with 2\,mm thick quartz windows to ensure
homogeneous transmission for wavelengths down to approximately
160\,nm.

The collecting optics was optimized for maximum light intensity. A
quartz lens in front of the lamp was used to concentrate the UV-light
at the surface of the scintillator. The UV-light is absorbed in the
first $\mu$m to mm of the liquid. Therefore, the emission takes place
also in the foremost layer of the scintillator sample.  For each
measurement, a light guide was used to collect the emitted light and
to pass it into the spectrometer's entrance slit. Looking at the
emitted light from the same side where the deuterium lamp illuminates
the sample ensures that propagation effects are almost negligible as
the emission takes place in the first few molecular planes.

\begin{figure}[h]
  \begin{center}
    \includegraphics[width=0.42\textwidth]{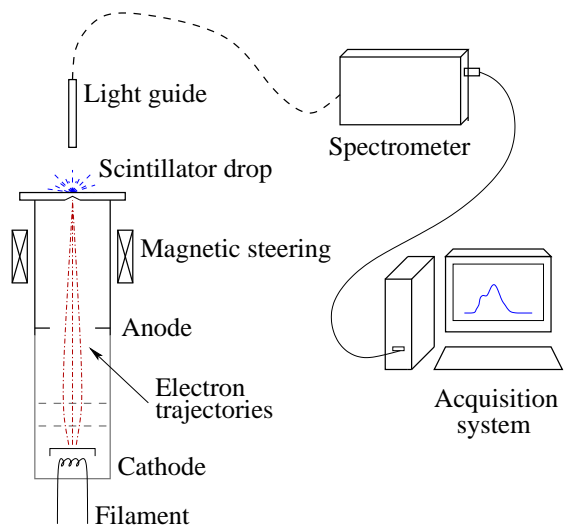}
  \end{center}
  \caption[SetUp_Egun_Diagram]{Setup for the spectral measurements of
    liquid scintillators when exciting the sample with $\sim$10\,keV
    electrons.\label{SetUp_Egun_Diagram}}
\end{figure}

Next, spectra of the scintillator mixtures were recorded after energy
deposition of charged particles. The radiation source used for these
measurements was a compact ($\sim$20\,cm long) low-energy
electron-beam device which is described in detail
in\,\cite{Wieser97}\cite{Morozov05}\cite{Ulrich09}. This device 
is housed in a glass-metal chamber which is kept at a vacuum of
$5\cdot10^{-7}$\,hPa. As indicated in
figure\,\ref{SetUp_Egun_Diagram}, a heated filament is used to produce
the electrons. Applying a DC voltage of 10-20\,kV, the electrons are
accelerated towards the exit window. Magnetic steering is used to
guide the beam in the $x$- and $y$- directions, perpendicular to the
electron trajectories. On the other end of the device, an exit window
is realized as a 300\,$\mu$m thick silicon wafer with an area of $0.7
\times 0.7$\,mm$^2$ and a 300\,nm thin foil in the center. The foil is
made of Si$_3$N$_4$ with an additional SiO$_2$ layer to reduce its 
internal mechanical stress. The energy loss in the
membrane depends on the electron energy: $\sim8$\% at 20\,keV,
$\sim15$\% at 15\,keV and $\sim40$\% at
10\,keV\,\cite{Morozov05}\cite{Morozov07}.

\begin{figure}[h]
  \begin{center}
    \includegraphics[angle=-90, width=0.37\textwidth]{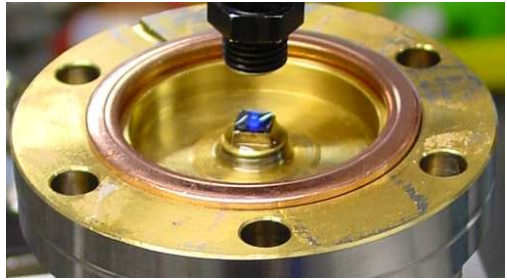}
  \end{center}
  \caption{Photograph of the scintillator drop on top of the exit
  window of the electron gun.\label{Egun_Blue}}
\end{figure}

Figure\,\ref{SetUp_Egun_Diagram} shows a schematic drawing of the
experimental setup.
The electron source was placed in a vertical position leading to
electrons exiting in upward direction. To observe the scintillation
light, a drop of the liquid sample was deposited on top of the
Si$_3$N$_4$ window. The drop covered a surface of about
5\,mm\,$\times$5\,mm and was about 1\,mm in
height. Figure\,\ref{Egun_Blue} shows a photograph of the blue-shining
scintillator drop on top of the Si$_3$N$_4$ window.
The end of the light guide was positioned about 1\,cm above the
window. The samples were taken from bottles with nitrogen atmosphere
and the measurements were performed immediately thereafter (within the
next 10\,min) to minimize oxygen diffusion into the sample, i.e. to
prevent a degradation of the scintillator. During the measurements, a
decrease of the emitted light intensity with time was observed. Two
probable explanations are favored: Either the intense $e^-$-radiation
damaged the liquid by destroying the constituent $\sigma$-bonds or the
scintillator drop evaporated partially with time, causing a loss of
scintillating material in the sample.

\section{Results and discussion}

Four samples were investigated:
\begin{itemize}
\item PXE with 2\,g/$\ell$ PPO,
\item PXE with 2\,g/$\ell$ PPO and 20\,mg/$\ell$ bisMSB,
\item LAB (P550 Q) with 2\,g/$\ell$ PPO ~and
\item PXE with 2\,g/$\ell$ PMP.
\end{itemize}
The solvents used were \emph{phenyl-o-xylethane} (PXE) and
\emph{linear-alkyl-benzene} (LAB, from Petresa Company type P550
Q\,\cite{Petresa}). The solutes dissolved were 
\emph{2,5-diphenyl-oxazole} (PPO),
\emph{1,4-bis-(o-methylstyryl)-benzene} (bisMSB) and
\emph{1-phenyl-3-mesityl-2-pyrazoline} (PMP).

\begin{figure*}[t]
  \includegraphics[width=0.99\textwidth]{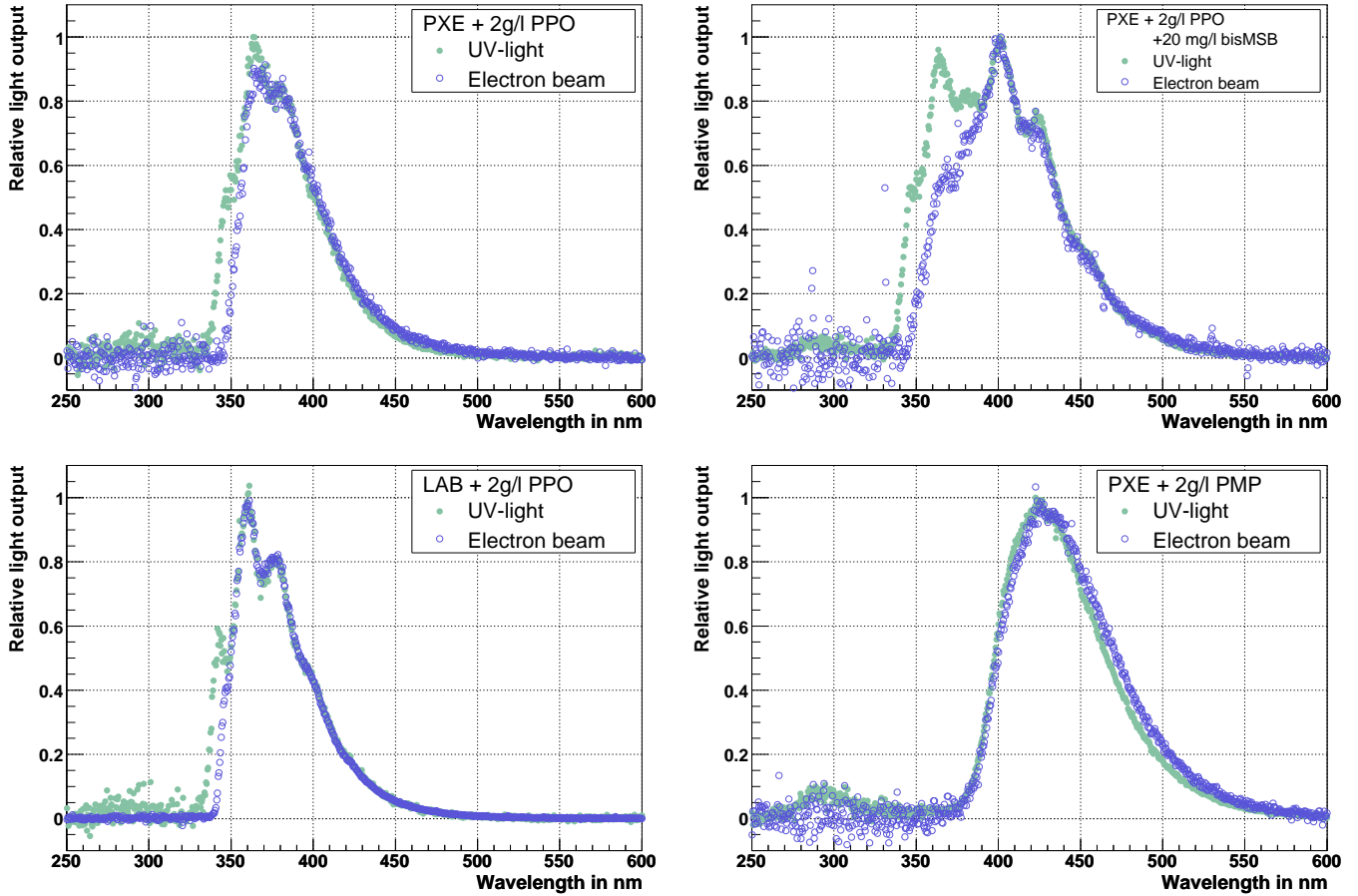}
  \caption[Emission from electron gun]{Spectra obtained with the
    electron-beam excitation method. The blue data points (open
    circles) represent the emission spectra of different scintillator
    mixtures after the light has propagated through the drop. For
    comparison\,\cite{PhD_TMarrodan}, data from the UV-light
    excitation method is also plotted (green full
    circles). Samples: PXE + 2\,g/$\ell$ PPO, PXE + 2\,g/$\ell$ PPO 
    + 20\,mg/$\ell$
    bisMSB, LAB + 2\,g/$\ell$ PPO, and PXE + 2\,g/$\ell$ PMP. The
    excitation is produced by $\sim$10\,keV electrons.\label{EmEgun_Spec}}
\end{figure*}

The spectra obtained by the methods described in section\,\ref{setups}
are shown in figure\,\ref{EmEgun_Spec}.
In each plot, the blue data points (open circles) represent the
emission spectra obtained by electron-beam excitation after the light
has propagated through the scintillator drop. The excitation was
produced by $\sim$10\,keV electrons. The data points obtained by
UV-light excitation are also plotted (green full circles) for
comparison\,\cite{PhD_TMarrodan}. The spectra are normalized in
intensity such that the highest intensity corresponds to 1. The
spectrometer was calibrated as described in section\,\ref{Spectrometer}. 
In the UV-region, the data points are more scattered; this is due to the
comparatively low sensitivity of the spectrometer in this wavelength region 
(see figure\,\ref{SpectrometerCorrCurv}).

Except for the PMP solute, the emission spectra show clear peaks. For
example in the case of the PPO solute (upper left in
figure\,\ref{EmEgun_Spec}), the peaks are approximately at 350, 365,
380 and 395\,nm. These peaks show the energy spacing between the first
singlet excited state, $S_1$, and the different vibrational levels of
the ground state of the solute, $S_{0i}$. The relative intensities of the
vibrational lines in the spectrum of PPO are compatible with the
literature results\,\cite{Berlman71}. The energy spacing between the
ground state and the first excited triplet state is in general
smaller, i.e. at longer wavelengths, than that of the first excited 
singlet state. 
As no significant contribution appears above $\sim 500$\,nm, it can 
be concluded that there is no evidence for
phosphorescence in the measured spectra. PMP displays a broad and less 
structured fluorescence spectrum which has also been reported in the
literature\,\cite{Guesten1978}.

For mixtures containing the solute PPO and the solvents PXE and LAB,
the shapes of the spectra above 375\,nm obtained by electron
excitation are almost identical to those measured by observing the
UV-illuminated front layer of the scintillator. The main difference
arises from the absorption at short wavelengths due to the
transmission through the scintillator drop. Short wavelengths are
rapidly absorbed within the first mm of the drop in the electron-beam 
experiment. 

The spectra of PXE with PMP differ only in the region from $\sim450$ 
to $\sim550$\,nm for the two excitation methods studied. The UV-induced 
intensity is maximally 5\% lower in this regime. This observation could be 
an indication of different molecular processes for the two studied excitation
methods. However, the effect is not significant enough to extract
relevant conclusions.

\begin{figure}[h]
  \begin{center}
    \includegraphics[width=0.49\textwidth]{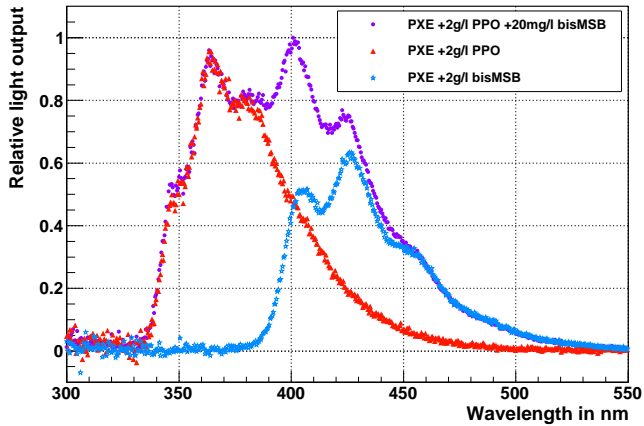}
    \caption[Emission and transmission spectra]{Zoom into the spectra
      of 2\,g/$\ell$ PPO and 20mg/$\ell$ bisMSB (violet points),
      2\,g/$\ell$ PPO (red triangles) and 2g/$\ell$ bisMSB
      (blue stars). All individual mixtures are dissolved in the solvent 
      PXE and excited by UV light. The
      three-component scintillator (violet points) can be seen to be the 
      sum of the spectra of
      both solutes. \label{EmTran_pxe2ppo20bisMSB_ppo_bisMSB}}
  \end{center}
\end{figure}

The spectrum of the mixture of PXE with 2\,g/$\ell$ PPO and
20\,mg/$\ell$ bisMSB (upper right in figure\,\ref{EmEgun_Spec}) shows very
interesting features. To detemine the origin of the various peaks, 
in figure\,\ref{EmTran_pxe2ppo20bisMSB_ppo_bisMSB} the spectra of PXE
with 2\,g/$\ell$ PPO and 20\,mg/$\ell$ bisMSB (violet points), PXE with
2\,g/$\ell$ PPO (red triangles) and PXE with 2\,g/$\ell$ bisMSB
(blue stars) are displayed together for the wavelength region between 
300\,nm and 550\,nm. These spectra have been obtained by 
exciting the scintillators with UV-light.
Figure\,\ref{EmTran_pxe2ppo20bisMSB_ppo_bisMSB} clearly shows that the 
three-component spectrum (violet points) results as
the sum of the PPO and bisMSB spectra. Although the concentration of
bisMSB is only 20\,mg/$\ell$, its contribution to the sum
spectrum is highly significant. 

The features of the spectrum in figure\,\ref{EmEgun_Spec} (upper right) 
can be explained by the following processes: When the UV-light from the 
deuterium lamp enters the scintillator, mostly PXE molecules are excited. 
The energy is rapidly transferred to PPO, mainly by nonradiative processes. 
Part of the energy in the PPO molecule is transferred further on to bisMSB
molecules in the same manner. As the concentration of bisMSB is low
(in the order of several mg/$\ell$), some of the PPO molecules find no
bisMSB partners. As a consequence, the light is partly emitted by the
PPO molecules themselves. The spectrum is therefore an overlap of the
PPO and the bisMSB spectra. Although the concentration of bisMSB is
low, the contribution of bisMSB to the sum spectrum is $\sim$\,40\%. 
This illustrates the efficient energy transfer from PPO to
bisMSB. Above 400\,nm, the two spectra obtained by UV-light or 
electron-beam excitation (see figure\,\ref{EmEgun_Spec}) are 
practically identical. Around 350\,nm, the absorption at short wavelengths 
can be observed in the spectrum with electron excitation as in the previous 
cases. The most interesting region is between 350 and 400\,nm, where the 
contribution of PPO to the spectrum by electron excitation is strongly 
reduced  compared to the spectrum obtained by UV-excitation. This can 
be explained  by the fact, that the light has to propagate through the 
scintillator drop (about a mm). Within this propagation distance, bisMSB 
very efficiently absorbs the light emitted by PPO. Consequently, the part 
of the spectrum due to the bisMSB emission is increased compared to that 
due to PPO.

Considering the differences in the spectra of PPO in the solvents PXE
and LAB (the two spectra on the left side of figure\,\ref{EmEgun_Spec}), 
a shift of $5-10$\,nm to longer wavelengths of all the PPO
vibrational peaks for the PXE solvent is observed. Furthermore, the
resolution of the vibrational levels differs: the peaks in the mixture
with LAB are better resolved. These effects can be explained by
considering the structure of the solvent molecules.  PXE
consists of two benzene rings and several CH$_3$ groups while LAB has 
only one benzene ring and a long hydrocarbon-chain. The solvent
molecules surrounding the benzene rings of PPO produce a mean
effective potential which influences the space available to the orbitals 
of the $\pi$-bonds of the solute molecule. The solute feels this
effective potential which might result in a different emission
spectrum. In addition, this mean potential can change the intrinsic
lifetime of the solute fluorescence\,\cite{Berlman71}. For example, it
has been found that the lifetime of PPO dissolved in PXE is shorter
than when dissolved in LAB\,\cite{Marrodan_Fl09}. Similar effects have
indeed been observed and described in the literature\,\cite{Berlman71}. 
A further consequence of this effective potential could 
be the wavelength difference of the peak emission of the solvents PXE 
and LAB, namely at 290 and 283\,nm, 
respectively, reported in\,\cite{CTF}\cite{SpecLAB_Buck}. In some of the 
spectra, the emission of the solvent can be seen around 300\,nm. The 
contribution is however, very small because the solute very effectively 
gets the energy transferred from the solvent to the solute.

\section{Conclusions}

Spectroscopy on organic liquid scintillators relevant for the proposed
LENA detector has been performed. The scintillators have been excited 
using a new method: a table-top electron beam source. 
For comparison, spectra are shown 
where the excitation occurs by UV-light. Apart from absorption effects due
to the propagation of light through a thin layer of scintillator, both
excitation methods return comparable results for the measured
samples. A precise know\-ledge of the emission spectra of liquid
scintillators is of great interest for the proposed LENA  detector 
as the propagation of the light (absorption and scattering processes) 
is wavelength dependent.

By studying the mixture of the solvent PXE with 2\,g/$\ell$ PPO and a
low concentration of bisMSB (20\,mg/$\ell$) it has been found that the
energy transfer between both wavelength-shifters happens very
efficiently already within short distances ($<$ mm). The electron-beam 
setup can be used in the future to measure emission spectra of further 
organic scintillators.

\section*{Acknowledgments}

We want to thank A. Morozov for his help during the measurements. This
work has been supported by funds of the Maier-Leibnitz-Laboratorium
(Munich), the Deutsche Forschungsgemeinschaft (DFG) (transregio TR27:
Neutrinos and Beyond) and the Munich Cluster of Excellence (Origin and
Structure of the Universe).

\end{document}